# Controlled atmosphere electrospinning of organic nanofibers with improved light emission and waveguiding properties

*Vito Fasano,[†] Maria Moffa,[§] Andrea Camposeo,[§] Luana Persano,[§] and Dario Pisignano[†, §]*

[†] Dipartimento di Matematica e Fisica "Ennio De Giorgi", Università del Salento, via Arnesano, I-73100, Lecce, Italy.

[§] Istituto Nanoscienze-CNR, Euromediterranean Center for Nanomaterial Modelling and Technology (ECMT), via Arnesano, I-73100, Lecce, Italy.

**ABSTRACT**

Electrospinning in controlled nitrogen atmosphere is developed for the realization of active polymer nanofibers. Fibers electrospun under controlled atmospheric conditions are found to be smoother and more uniform than samples realized by conventional electrospinning processes performed in air. In addition, they exhibit peculiar composition, incorporating a greatly reduced oxygen content during manufacturing, which favors enhanced optical properties and increases emission quantum yield. Active waveguides with optical losses coefficients lowered by ten times with respect to fibers spun in air are demonstrated through this method. These findings make the process very promising for the highly-controlled production of active polymer nanostructures for photonics, electronics and sensing.

The interest to active polymer nanostructures for photonics and optoelectronics has vastly increased in the last decade, due to their photophysical properties, high flexibility, and low-cost manufacturing processes.[1-4] Methods for obtaining conjugated polymer nanorods and fibers are various and versatile, comprising self-assembly,[5-7] synthesis in porous templates,[1,8] interfacial polymerization,[9] and electrospinning.[10-14] Demonstrated devices embedding polymer nanowires include organic light-emitting diodes,[15,16] chemical sensors,[2,17] solid-state lasers,[1,18] photovoltaic cells,[19] and field effect transistors.[20,21] In all these architectures and in the related processing steps, special care has to be paid to avoid degradation of the material physico-chemical properties, which is likely to occur due to oxidation.[22-24] For instance, photo-oxidation leads to significant variations in the emission spectra and to a large reduction of the luminescence quantum yield of conjugated polymers.[25-27] These degradation pathways have been analyzed by photoabsorption,[24] vibrational spectroscopy and photoluminescence decay dynamics [25,28] and near-field scanning optical microscopy.[22]

In this framework, most of studies have investigated how to improve the stability of active organics during device operation, particularly through encapsulation strategies,[29] while lower attention has been focused on the correlation between the processing conditions and the resulting morphological and optical features of the polymer nanostructures. In particular, it is noteworthy that a-posteriori device encapsulation, while limiting oxygen diffusion towards active polymer elements during device life, might be unsuited to efficiently block contaminations embedded in nanostructures just during manufacturing. To this aim, methods of preparation working under controlled atmosphere should be designed and developed. In addition, realizing polymer nanowires and nanofibers under controlled atmosphere conditions might impact on various morphological properties, particularly on surface roughness which directly relates with the evaporation rates of the solvents used during processing, and which in turn affects optical properties and light scattering from the nanostructures.

Electrospinning is especially interesting in this respect. This technology is cost-effective and versatile, and allows fibers to be realized with intrinsic molecular orientation along their longitudinal axis, due to solution jet stretching by an applied electric field.[30,31] Fibers are obtained following solvent evaporation from the jet, which makes evaporation rates and fluid-atmosphere interactions very important in affecting the overall process outcome and the morphology of deposited filaments.[32-35] Electrospun fibers embedding conjugated polymers show bright light emission and significant waveguiding,[13,36] however when realized in ambient conditions they are unavoidably loaded with some amounts of oxygen and moisture incorporated by the jets. For these reasons, developing the production of these materials under controlled atmosphere is especially worthwhile, since it has great potential for leading to fibers with improved surface morphology and enhanced optical properties. Interestingly, while a few equipment are already commercialized which allow electrospinning to be carried out in a protected environment,[37,38] the impact of processes performed under controlled atmosphere on nanofiber properties is basically unexplored.

Here we report for the first time on active electrospun fibers realized under nitrogen atmosphere conditions. Fibers are made of the light-emitting conjugated polymers, poly[(9,9-dioctylfluorenyl-2,7-diyl)-*alt*-co-(1,4-benzo-[2,1',3]-thiadiazole)] (F8BT) and poly[2-methoxy-5-(2-ethylhexyloxy)-1,4-phenylenevinylene] (MEH-PPV), and exhibit peculiar morphological and compositional properties, greatly reduced surface roughness, reduced oxygen incorporation, and improved emission and waveguiding. The set-up developed for electrospinning in nitrogen atmosphere is photographed in Fig. 1a. A glove-box is used, with suitable feedthroughs for electrical connections. Polymer solutions are delivered with controlled flux to the spinneret by a precision pump placed in the chamber, and the process is performed with oxygen below 2 ppm and humidity below 5 ppm, respectively. Both randomly oriented and aligned nanofibers can be deposited, by using collectors of suitable geometry in the processing chamber. Two exemplary mats made of light-emitting fibers are shown in Fig. 1b,c. For sake of comparison, electrospinning experiments in air are carried out in parallel. As typical of spinning conjugated polymer solutions, complex shapes are formed by the fluid at the spinneret, due to the competition of the quite high solvent evaporation rate, the mass delivery by the pump and the pulling electric field, and to the resulting interplay of local polymer accumulation points and Taylor cone[39] formation (Fig. 1d,e). The interaction with the surrounding atmosphere might lead to modifications of such shapes. For instance, imaging the onset of spinning suggests a more prompt jetting of MEH-PPV/polyvinylpyrrolidone (PVP) solutions in the controlled chamber, namely a lower material accumulation at the needle (Fig. 1d,e).

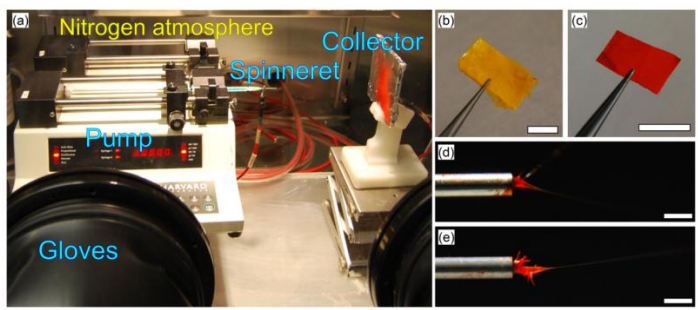

Figure 1. (a) Set-up for electrospinning in nitrogen atmosphere. Gloves and electrical feed-through connections allow the equipment to be handled and electrical wires to provide power supply to the set-up, respectively. (b, c) Light-emitting fibers mats made of F8BT (b) and MEH-PPV/PVP (c) electrospun in the controlled chamber. Scale bar = 1 cm. (d, e) Images of the onset of electrospinning of MEH-PPV/PVP solutions in the nitrogen atmosphere (d) and in air (e). Scale bar = 1 mm.

The resulting sample morphologies at microscale are shown in Fig. 2. Fibers spun in nitrogen and in air have comparable average diameters (1.9±0.5 μm *vs.* 1.7±0.4 μm for F8BT and 380±130 nm *vs.* 390±180 nm for MEH-PPV/PVP, respectively, Fig. S1). Instead, the process conditions considerably affect the surface of fibers. When spun at ambient conditions, F8BT fibers display a wrinkled morphology with dimple-shaped features, while fibers spun in controlled atmosphere exhibit a much smoother surface (insets of Fig. 2a,b). An analogous behavior is found in MEH-PPV/PVP fibers and ribbons (Fig. 2d,e).

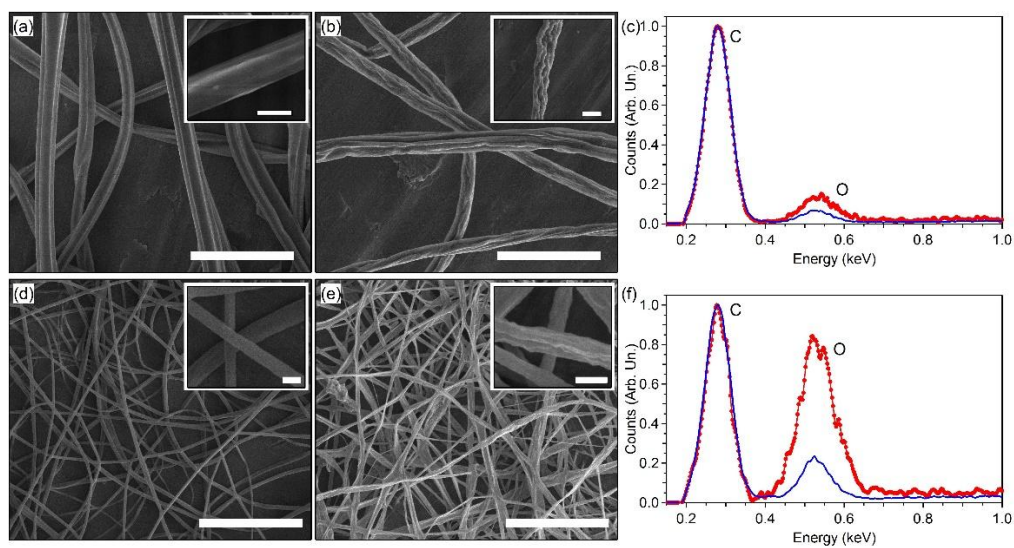

Figure 2. (a,b) SEM micrographs of F8BT fibers electrospun in controlled atmosphere (a) and in air (b). Scale bar = 10 μm. Insets: Close-up of individual fibers (scale bar = 2 μm). (c) EDS spectra of F8BT fibers realized in air (red circles) and in nitrogen (blue line). (d, e) SEM images of MEH-PPV/PVP fibers electrospun in nitrogen (d) and in air (e). Scale bar = 10 μm. Insets: Close-up of fibers (scale bar = 0.5 μm). (f) Corresponding EDS spectra for samples spun in air (red circles) and in nitrogen (blue line).

These findings are directly related to different humidity in the process environment. Indeed, it is established that, at sufficient atmospheric humidity, water condensates on filaments during electrospinning, thus priming phase separation, and pores as well as sub-μm surface features are formed upon evaporation of such water and of volatile solvents.[34,35] Therefore, carrying out the process under an atmosphere with very low humidity allows smoother electrospun fibers to be achieved, which would exhibit improved emission and waveguiding properties due to weaker light-scattering at their surface. Also, incorporating a reduced content of oxygen during fabrication is critically important to many applications in photonics and electronics. Energy-dispersive X-ray spectroscopy (EDS) data, normalized to the intensity of the carbon peak, highlight a drastic decrease of the overall oxygen amount in fibers spun under nitrogen (Fig. 2c,f). Indeed, the resulting fibers emit bright light (Fig. 3a,b), with either no significant spectral change compared to fibers spun in air (Fig. 3c), or with slightly red-shifted absorption ($\Delta\lambda_{Abs} = 20$ nm) and emission ($\Delta\lambda_{PL} = 10$ nm) lineshapes (Fig. 3d), in part attributable to the spectral behavior of morphology-dependent light scattering.[40] These spectral differences found in MEH-PPV/PVP fibers can be also correlated to reduced oxidation phenomena. For fibers spun in air, oxidation is promoted by the excess of incorporated oxygen, reducing the effective conjugation length of the polymer chains.[22,26] Analogously, absolute photoluminescence quantum yields (φ) for fibers spun in controlled atmosphere are generally higher than those of fibers spun in air (φ = 0.51 and 0.49 for F8BT, and φ = 0.06 and 0.03 for MEH-PPV/PVP, respectively). To investigate oxidation effects, we also collect the Fourier Transform Infrared (FTIR) spectra of fibers, repeating measurements after different intervals of UV irradiation. These experiments are performed placing samples under a $N_2$ flow, in order to minimize effects related to the diffusion of extra oxygen into the fibers during measurements. As shown in Fig. 4, MEH-PPV-PPV fibers present a peak characteristic of the PVP carrier (stretching vibration mode of the amide group at 1660 cm$^{-1}$),[41] as well as typical transitions of the conjugated polymer such as the ether C–O–C stretching (1040 cm$^{-1}$), the phenyl-oxygen stretching (1204 cm$^{-1}$), the phenyl ring modes (1500 cm$^{-1}$ and 1415 cm$^{-1}$) and the asymmetric CH$_2$ deformation (1465 cm$^{-1}$).[23,42] Following UV exposure, an increase of the peak at about 1730 cm$^{-1}$ (associated to carbonyl groups[28,42]) is measured for MEH-PPV/PVP fibers realized in air compared to those spun in controlled atmosphere (insets in Figure 4). Indeed, the carbonyl peak is known to be very sensitive to the photo-oxidation of the conjugated polymer.[28,42] FTIR spectra of F8BT fibers are instead more stable (Figure S2), consistently with the EDS results which indicate a less pronounced difference in the oxygen incorporated in standard and controlled-atmosphere processes (Figure 2c). Overall, FTIR results support quantum efficiency measurements, i.e. an increasing oxygen content embedded during electrospinning leading to enhanced degradation pathways in the light-emitting nanostructures.

Dishomogeneities along conjugated polymer nanofibers are indicative of the complex configurational processes of macromolecules during electrospinning,[43] and of the consequently different aggregation states of backbones in the solid state. These issues are evaluated by spectrally-resolved confocal maps, which allow the uniformity of emission from different regions of individual fibers to be assessed in detail. The spectra from various fiber segments at microscopic scale show differences below 10% in their 0-0 and 0-1 vibronic replica intensity ratios and widths (Figure 5a), which may be attributed to local variations in the formation of interchain species, aggregates and packing of the conjugated polymeric chains promoted by electrified jets.[26,40,44]

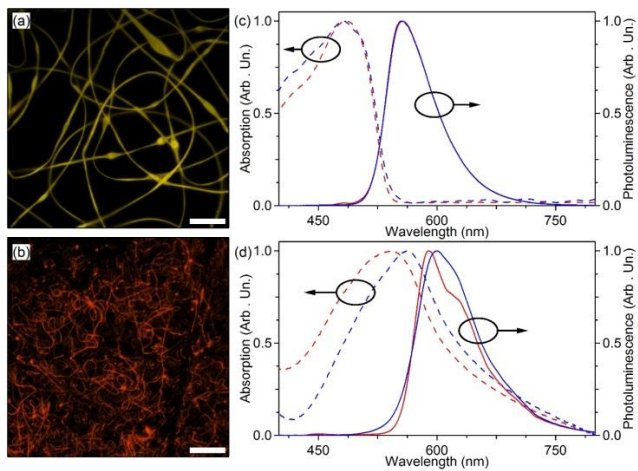

Figure 3. Fluorescence confocal micrographs of F8BT (a) and MEH-PPV/PVP (b) fibers electrospun in controlled atmosphere. Scale bars = 100 μm. Excitation wavelength $\lambda$ = 408 nm (a) and 488 nm (b). (c, d) Normalized absorption (dashed lines) and photoluminescence (continuous lines) spectra of F8BT (c) and MEH-PPV/PVP (d) fibers electrospun in controlled atmosphere (blue lines) and in air (red lines).

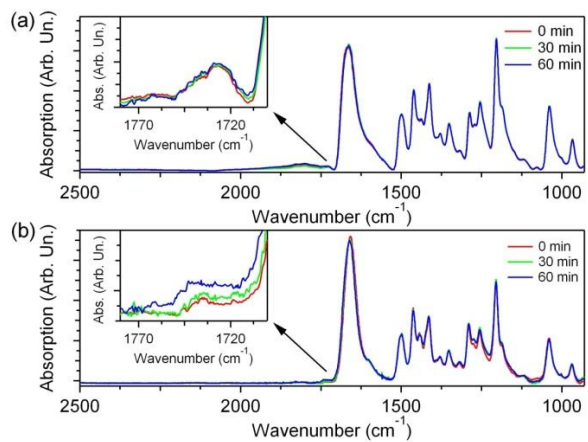

Figure 4. FTIR absorption spectra of MEH-PPV/PVP fibers realized in controlled atmosphere (a) and in air (b). Spectra are acquired before UV exposure (red lines) and after 30 minutes (green lines) and 60 minutes (blue lines) of UV irradiation, respectively. Insets: close-up of FTIR spectra around 1730 cm$^{-1}$.

Interestingly, fibers spun in air show a higher degree of dishomogeneity, with variations up to 15% in the intensity ratios and spectral widths of their vibronic transitions, measured in various points along the fiber longitudinal axis (Figures S3 and S4). μ-PL also allows the orientational properties of polymer molecules within the electrospun structures to be investigated. Spectra from single F8BT fibers, collected with a polarization filter either parallel ($PL_{\parallel}$) or perpendicular ($PL_{\perp}$) to the fiber axis, are displayed in the inset of Figure 5b, evidencing a polarized light emission. The distribution of the photoluminescence polarization ratio ($\chi = PL_{\parallel} / PL_{\perp}$) measured on 70 fibers indicates that, on average, $PL_{\parallel}$ is about three times more intense than $PL_{\perp}$ (Fig. 5b). The preferential alignment of chromophore dipoles is also highlighted in Fig. 5c, where we show the single fiber emission intensity as a function of the angle (θ) defined by the polarization of the emitted light and the longitudinal fiber axis, together with the best fit by a Malus law [~$\cos^2(\theta)$]. Similar findings are obtained for fibers spun in air (Figure S5), which indicates that average orientational properties of chromophores within light-emitting fibers, as induced by solution jet stretching,[36,45] are not significantly affected by the electrospinning atmosphere.

The stable emission, improved quantum yield and smoother surface morphology make light-emitting fibers electrospun in controlled atmosphere conditions very promising for realizing miniaturized photonic devices, particularly waveguides. To this aim, fibers are deposited on $MgF_2$ (refractive index, $n = 1.37$), minimizing light-coupling into the underlying substrate.[13,36] To investigate photon propagation losses along fibers, the photoluminescence escaping from a fiber tip is imaged and its intensity is measured as a function of the distance ($d$) from the excitation region (Fig. 6a,b). The measured intensity ($I$) shows a well-behaved exponential decay upon increasing $d$ (Fig. 6c), namely $I = I_0 \exp(-\alpha d)$, where $I_0$ is the intensity of the fiber-coupled emission, and α indicates the optical loss coefficient. A much lower optical loss coefficient is found for fibers spun in nitrogen (α ≅ 80 cm$^{-1}$, vs. α ≅ 800 cm$^{-1}$ for fibers spun in air).

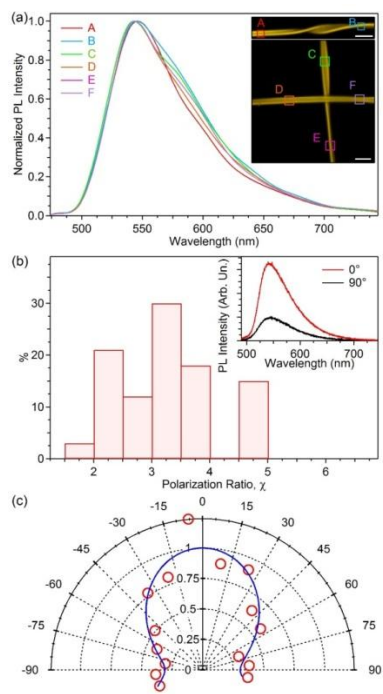

Figure 5. (a) Spatially-resolved photoluminescence spectra and corresponding confocal fluorescence images (insets) of F8BT fibers electrospun in controlled atmosphere. Excitation wavelength λ = 408 nm. Each emission spectrum, measured in a different region of the fibers (squares in the inset), is normalized to its maximum value. Inset: scale bar = 5 μm. The size of each analyzed square is 2.5×2.5 μm$^2$. (b) Distribution of emission polarization ratio, χ, for individual F8BT fibers. Inset: polarized emission spectra obtained with the analyzer axis parallel (red line) and perpendicular to (black line) the fiber length. (c) Polar plot of the normalized photoluminescence intensity (circles) as function of the angle of the analyzer filter axis, θ, measured with respect to the fiber longitudinal axis (θ = 0° for polarization filter parallel to the fiber axis, θ = 90° for polarization filter perpendicular to the fiber axis). Continuous line: best fit to data by a cos$^2$(θ) law.

As comparison, previously demonstrated nanofiber species show $\alpha$ values between 100 cm$^{-1}$ and 1000 cm$^{-1}$.[8,13,36] Typically, optical losses in these miniaturized, organic active waveguides are associated with self-absorption and with scattering from surface or bulk defects. In our case, the much lower losses in fibers spun in nitrogen is mainly related to the absence of light-scattering dishomogeneities along the fiber surface, which are instead present in fibers realized in air (white arrows in Figure 6b), as found by scanning electron microscopy (SEM). Indeed, surface scattering losses are particularly significant in the case of fibers with micrometric diameters, where high-order guided modes interact strongly with the waveguide surface.[46] These issues highlight the importance of obtaining smoother optical interfaces and active microsystems, as can be achieved by electrospinning in highly controlled atmospheric conditions. Lower light propagation losses and increased emission efficiency are advantageous for developing devices, optical sensors and low-threshold lasers based on nanofibers. For instance, being typically based on luminescence quenching by energy or charge transfer to an analyte,[2,47,48] optical sensors based on nanofibers with higher quantum yields would ultimately exhibit enhanced sensitivity due to the lower processing-related extrinsic doping. For laser applications,[1,3,18] increased quantum yields as well as decreased propagation losses would contribute to reduce excitation thresholds, because of the higher number of emitted photons available for light amplification by stimulated emission along nanofibers. In electronics, the availability of conductive polymer nanofibers with reduced incorporation of oxygen and moisture might lead to lower ambient doping and to a better saturation behavior of field-effect transistors.[49,50]

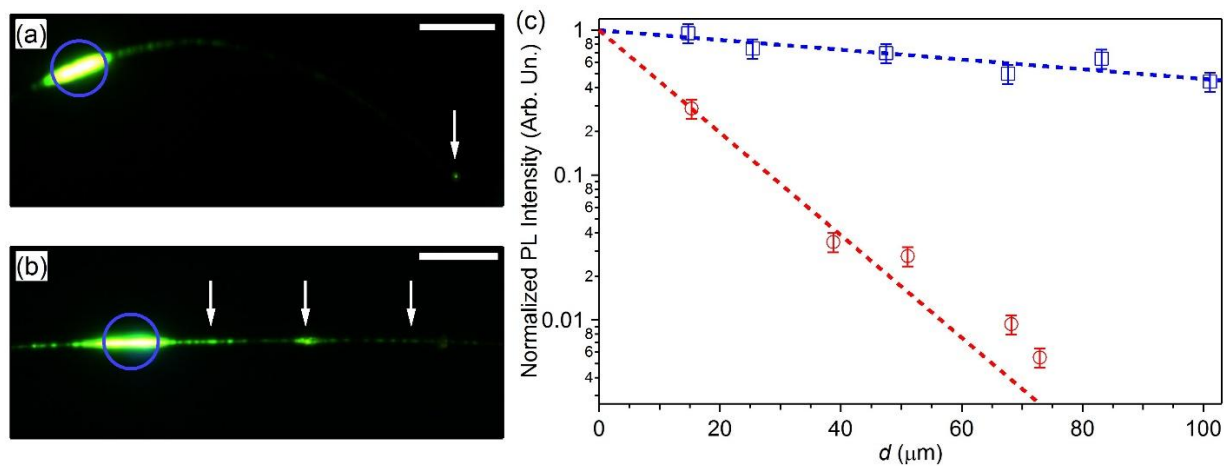

Figure 6. (a, b) Micrographs of individual F8BT fibers, spun in controlled atmosphere (a) and in air (b), and excited by a focused laser spot (blue circles). Scale bars = 20 μm. White arrows in (a) and (b) indicate the fiber tip and the defects and dishomogeneities along the fiber axis, respectively. (c) Spatial decay of the light intensity waveguided along a single fiber vs. distance, *d*, from the excited region. Squares (circles): fibers electrospun in controlled atmosphere (air). Dashed lines: best fits to exponential decays.

**CONCLUSIONS**

Uniform and bright light-emitting fibers based on conjugated polymers are realized by electrospinning in controlled atmospheric conditions. Electrospun fibers are critically sensitive to the process atmosphere, being smoother and more uniform when produced in nitrogen environment compared to samples spun in air. This effect is synergic with the reduced oxygen and moisture incorporation during electrospinning, and induces significantly enhanced optical properties and improved waveguiding performances of the resulting light-emitting fibers. These findings make the process highly interesting for the realization of improved active nanofibers for optical sensors, nanostructured light-emitting devices and lasers, and nanoelectronics.

**METHODS**

*Electrospinning.* Solutions of (i) F8BT ($M_w$ = 132 kDa, American Dye Source, Inc.), and (ii) MEH-PPV ($M_w$ = 150-250 kDa, Sigma-Aldrich Co.) with PVP ($M_w$ = 130 kDa, Alfa Aesar) are prepared by (i) tetrahydrofuran/dimethylsulfoxide at 9:1 (v:v) and by (ii) chloroform/dimethylsulfoxide at 9:1 (v:v) relative solvent concentration, respectively. MEH-PPV and PVP are blended at 1:1 (w:w) relative concentration, which is found to assure bright emission together with the formation of uniform fibers. Solutions are made with a total polymer concentration (i) 70 mg/mL and (ii) 30 mg/mL for F8BT and MEH-PPV/PVP, respectively. Solutions are stirred for 24 h before use in electrospinning processes. The electrospinning set-up consists of a 1 mL syringe connected to a pump (Harvard Apparatus) and a high-voltage power supply (Glassman Series EH) applying a 10 kV voltage bias to the syringe needle (21 gauge). Steady flows of 0.5 and 1.0 mL/h are provided by the pump for F8BT and MEH-PPV/PVP solutions, respectively. F8BT and MEH-PPV/PVP fibers are collected on grounded Al sheets or microscopy slides placed at 15 cm from the needle tip. Identical process parameters are used for experiments performed in controlled atmosphere and in in air. Spinning experiments in nitrogen atmosphere are performed in a glove box (Jacomex, GP[Concept]) equipped with a $O_2$ galvanic cell oxygen sensor with ± 0.1 ppm resolution and ± 1 ppm accuracy and a $H_2O$ ceramic sensor (accuracy ± 2°C dew point). During the electrospinning experiments and from run to run, we measured $O_2 \leq 2$ ppm and $H_2O < 5$ ppm with typical fluctuations of ± 0.2 ppm  and ± 0.1 ppm, respectively. Reference experiments in air are carried out with temperature and humidity values of 22-24 °C and 30-40%, respectively.

*Morphological and confocal characterization.* Fibers are coated with Cr and inspected by SEM (FEI Nova NanoSEM 450) at 10 kV. Elemental analysis is performed by EDS, with acceleration voltage 15 kV. Confocal fluorescence imaging is carried out by a laser-scanning microscope (Nikon A1R-MP) equipped with a spectral scan head (Nikon). The confocal system is composed by an inverted microscope (Eclipse Ti, Nikon), a $20\times$ objective (numerical aperture $N.A. = 0.50$) and a set of laser sources ($\lambda = 408$ nm and 488 nm). The sample emission is collected through the microscope objective, and the intensity is analyzed by the spectral detection scan head equipped with a multi-anode photomultiplier. During electrospinning, the early-stage of jetting is imaged by a photography set-up composed of a reflex camera (Nikon D40x, Nikon Corp., Japan) equipped with a fixed focal length micro-objective (200 mm, f/4, Nikkor, Japan) and a macro extension tubes kit. The whole optical system is mounted on a photography tripod used to minimize the motion blur effect on the acquired images.

*Spectroscopy.* Absorption spectra of fibers are collected by a double beam ultraviolet-visible spectrophotometer and by a FTIR spectrometer (Perkin Elmer). Infrared spectra are recorded under a controlled $N_2$ flow with a 1 cm$^{-1}$ resolution, averaged over 50 scans and baseline-corrected, both before and after exposure to UV light, which is carried out by an 8 W lamp (Spectroline, EN-180L/FE, Spectronics, $\lambda = 365$ nm). Photoluminescence spectra are measured exciting samples by a continuous wave (cw) diode laser ($\lambda = 405$ nm, μLS Micro Laser Systems, Inc.) and collecting the emission by a fiber-coupled spectrometer (USB 4000, Ocean Optics). The absolute quantum efficiency (φ) of fibers is determined by an integrating sphere (Labsphere), exciting by a cw diode laser and analyzing photoluminescence by a spectrometer. All the spectra are corrected for the apparatus spectral response (integrating sphere, optical fiber and spectrometer). The polarization properties of individual fibers are investigated by micro-photoluminescence (μ-PL), using an inverted microscope (IX71, Olympus). A laser beam is focused onto single fibers through the microscope objective ($N.A. = 0.5$, spot diameter of a few microns) for photo-excitation. The polarization of the excitation laser is parallel with the fiber longitudinal axis. A rotating polarizer is used for characterizing the photoluminescence polarization state of the sample, whose emission is coupled into an optical fiber and spectrally analyzed.

*Waveguiding.* Fiber waveguiding properties are also studied by μ-PL, using a Peltier cooled charge-coupled device (Leica, DFC 490). The beam from a cw diode laser is focused on samples through a dichroic mirror and the microscope objective (*N.A.* = 0.5). Part of the light emitted by the excited nanofiber region is coupled into the polymer wire and then waveguided. The optical loss coefficient is obtained analyzing the decay of the light intensity collected from the fiber tip, as a function of the distance from the exciting laser spot.

## AUTHOR INFORMATION

**Corresponding Author**

Dario Pisignano. E-mail address: dario.pisignano@unisalento.it

## ASSOCIATED CONTENT

**Supporting Information.**

Diameter distributions and further fiber characterization material is available free of charge via the Internet at http://pubs.acs.org.

## ACKNOWLEDGMENT

The research leading to these results has received funding from the European Research Council under the European Union's Seventh Framework Programme (FP/2007-2013)/ERC Grant Agreement n. 306357 (ERC Starting Grant "NANO-JETS"). The Apulia Networks of Public Research Laboratories Wafitech (09) and M. I. T. T. (13) are also acknowledged. M. Coviello is acknowledged for assistance in fabrication experiments.

## REFERENCES

1    O'Carroll, D.; Lieberwirth, I.; Redmond, G. *Nat. Nanotechnol.* **2007**, *2*, 180.

2    Kim, F. S.; Ren, G.; Jenekhe, S. A. *Chem. Mater.* **2011**, *23,* 682.

3    Camposeo, A.; Persano, L.; Pisignano, D. *Macromol. Mater. Eng*. **2013**, *298*, 487.

4    Persano, L.; Camposeo, A.; Pisignano, D. *Prog. Polym. Sci.* **2015**, *43*, 48.

5        Liu, J.; Sheina, E.; Kowalewski, T.; McCullough, R. D. *Angew. Chem. Int. Ed.* **2002**, *41,* 329.

6        Samitsu, S.; Shimomura, T.; Heike, S.; Hashizume, T.; Ito, K. *Macromolecules* **2008**, *41,* 8000.

7        Wang, S.; Kappl, M.; Liebewirth, I.; Müller, M.; Kirchhoff, K.; Pisula, W.; Müllen, K. *Adv. Mater.* **2012**, *24*, 417.

8        O'Carroll, D.; Lieberwirth, I.; Redmond, G. *Small* **2007**, *3,* 1178.

9        Huang, J.; Virji, S.; Weiller, B. H.; Kaner, R. B. *J. Am. Chem. Soc.* **2003**, *125*, 314.

10       Reneker, D. H.; Chun, I. *Nanotechnology* **1996**, *7,* 216.

11       Greiner, A.; Wendorff, J. H. *Angew. Chem. Int. Ed.* **2007,** *46,* 5670.

12       Li, D.; Xia, Y. *Adv. Mater.* **2004**, *16,* 1151.

13       Di Benedetto, F.; Camposeo, A.; Pagliara, S.; Mele, E.; Persano, L.; Stabile, R.; Cingolani, R.; Pisignano, D. *Nat. Nanotechnol.* **2008**, *3,* 614.

14       Pisignano, D. Polymer Nanofibers, Royal Society of Chemistry, Cambridge, **2013**.

15       O'Carroll, D.; Iacopino, D.; O'Riordan, A.; Lovera, P.; O'Connor, É.; O'Brien, G. A.; Redmond G. *Adv. Mater.* **2007**, *20*, 42.

16       Vohra, V.; Giovanella, U.; Tubino, R.; Murata, H.; Botta, C. *ACS Nano* **2011**, *5*, 5572.

17       Liu, H.; Kameoka, J.; Czaplewski, D. A.; Craighead, H. G. *Nano Lett*. **2004**, *4*, 671.

18    Persano, L.; Camposeo, A.; Del Carro, P.; Fasano, V.; Moffa, M.; Manco, R.; D'Agostino, S.; Pisignano, D. *Adv. Mater.* **2014**, *26*, 6542.

19    Wu, P.-T.; Xin, H.; Kim, S. F.; Ren, G.; Jenekhe, S. A. *Macromolecules* **2009**, *42*, 8817.

20    Babel, A.; Li, D.; Xia, Y.; Jenekhe, S. A. *Macromolecules* **2005**, *38*, 4705.

21    Lee, S. W.; Lee, H. J.; Choi, J. H.; Koh, W. G.; Myoung, J. M.; Hur, J. H.; Park, J. J.; Cho, J. H.; Jeong, U. *Nano Lett.* **2010**, *10,* 347.

22    Credo, G. M.; Lowman, G. M.; DeAro, J. A.; Carson, P. J.; Winn, D. L.; Buratto, S. K. *J. Chem. Phys.* **2000**, *112*, 7864.

23    Wang, Y.; Peng, Y.; Mo, Y.; Yang, Y; Zheng X. *Appl. Phys. Lett.* **2008**, *93,* 231902.

24    Sutherland, D. G. J.; Carlisle, J. A.; Elliker, P.; Fox, G.; Hagler, T. W.; Jimenez, I.; Lee, H. W.; Pakbaz, K.; Terminello, L. J.; Williams, S. C.; Himpsel, F. J.; Shuh, D. K.; Tong, W. M.; Jia, J. J.; Callcott, T. A.; Ederer, D. L. *Appl. Phys. Lett.* **1996**, *68,* 2046.

25    Yan, M.; Rothberg, L. J.; Papadimitrakopoulos, F.; Galvin, M. E.; Miller, T. M. *Phys. Rev. Lett.* **1994**, *73*, 744,

26    Nguyen, T.-Q.; Martini, I. B.; Liu, J.; Schwartz, B. J. *J. Phys. Chem. B* **2000**, *104*, 237.

27    Wang, Y.-H.; Peng, Y.-J.; Mo, Y.-Q.; Yang, Y.-Q.; Zheng, X.-X. *Appl. Phys. Lett.* **2008**, *93*, 231902.

28    Golovnin, I. V.; Bakulin, A. A.; Zapunidy, S. A.; Nechvolodova, E. M.; Paraschuk D. Y. *Appl. Phys. Lett.* **2008**, *92,* 243311.

29      Park, J.-S.; Chae, H.; Chung, H. K.; Lee, S. I. *Semicond. Sci. Technol*. **2011**, *26*, 034001.

30      Kakade, M. V.; Givens, S.; Gardner, K.; Lee, K. H.; Chase, B.; Rabolt, J. F. *J. Am. Chem. Soc.* **2007**, *129*, 2777.

31      Pagliara, S.; Vitiello, M. S.; Camposeo, A.; Polini, A.; Cingolani, R.; Scamarcio, G.; Pisignano, D. *J. Phys. Chem. C* **2011**, *115,* 20399.

32      Bognitzki, M.; Czado, W.; Frese, T.; Schaper, A.; Hellwig, M.; Steinhart, M.; Greiner, A.; Wendorff, J. H. *Adv. Mater.* **2001**, *13*, 70.

33      Koombhongse, S.; Liu, W.; Reneker, D. H. *J. Polym. Sci. Part B: Polym. Phys.* **2001**, *39*, 2598.

34      Megelski, S.; Stephens, J. S.; Chase, D. B.; Rabolt, J. F. *Macromolecules* **2002**, *35*, 8456.

35      Casper, C. L.; Stephens, J. S.; Tassi, N. G.; Chase, D. B.; Rabolt, J. F. *Macromolecules* **2004**, *37*, 573.

36      Fasano, V.; Polini, A.; Morello, G.; Moffa, M.; Camposeo, A.; Pisignano, D. *Macromolecules* **2013**, *46*, 5935.

37      http://www.imetechnologies.nl/file/970/IME_Electrospinning_equipment_2015.pdf

38      http://www.electrospunra.com/sub/es210.html#page_1/

39      Yarin, A. L.; Koombhongse, S.; Reneker, D. H. *J. Appl. Phys*. **2001**, *90*, 4836.

40      Ishii, Y.; Murata, H. *J. Mater. Chem.* **2012**, *22*, 4695.

41      Zheng, S.; Guo, Q.; Mi, Y. *J. Polym. Sci. Pol. Phys.* **1999**, *37*, 2412.

42      Cumpston, B. H.; Jensen, K. F. *J. Appl. Polym. Sci.* **1998**, *69*, 2451.

43      Camposeo, A.; Greenfeld, I.; Tantussi, F.; Pagliara, S.; Moffa, M.; Fuso, F.; Allegrini, M., Zussman, E.; Pisignano, D. *Nano Lett.* **2013**, *13*, 5056.

44      Li, D.; Babel, A.; Jenekhe, S. A.; Xia, Y. *Adv. Mater.* **2004**, *16,* 2062.

45      Richard-Lacroix, M.; Pellerin, C. *Macromolecules* **2013**, *46*, 9473.

46      Hunsperger, R. G. *Integrated Optics, Theory and Technology*, Springer-Verlag, New York, **1982**.

47      Wang, X.; Drew, C.; Lee, S.-H.; Senecal, K. J.; Kumar, J.; Samuelson, L. A. *Nano Lett.* **2002**, *2*, 1273.

48      Camposeo, A.; Moffa, M.; Persano, L. Electrospun Fluorescent Nanofibers and Their Application in Optical Sensing. In *Electrospun nanofibers for high performance sensors*, Macagnano, A.; Zampetti, E.; Kny, E.; Eds.; Springer-Verlag: Berlin Heidelberg, **2015**. pp. 129-155.

49      Abdou, M. S. A.; Orfino, F. P.; Son, Y.; Holdcroft. *J. Am. Chem. Soc.* **1997**, *119,* 4518.

50      Hoshino, S.; Yoshida, M.; Uemura, S.; Kodzasa, T.; Takada, N.; Kamata, T.; Yase, K. *J. Appl. Phys.* **2004**, 95, 5088.

# SUPPORTING INFORMATION

# Controlled atmosphere electrospinning of organic nanofibers with improved light emission and waveguiding properties

*Vito Fasano,[†] Maria Moffa,[§] Andrea Camposeo,[§] Luana Persano,[§] and Dario Pisignano[†,§]*

[†] Dipartimento di Matematica e Fisica "Ennio De Giorgi", Università del Salento, via Arnesano, I-73100, Lecce, Italy.

[§] Istituto Nanoscienze-CNR, Euromediterranean Center for Nanomaterial Modelling and Technology (ECMT), via Arnesano, I-73100, Lecce, Italy.

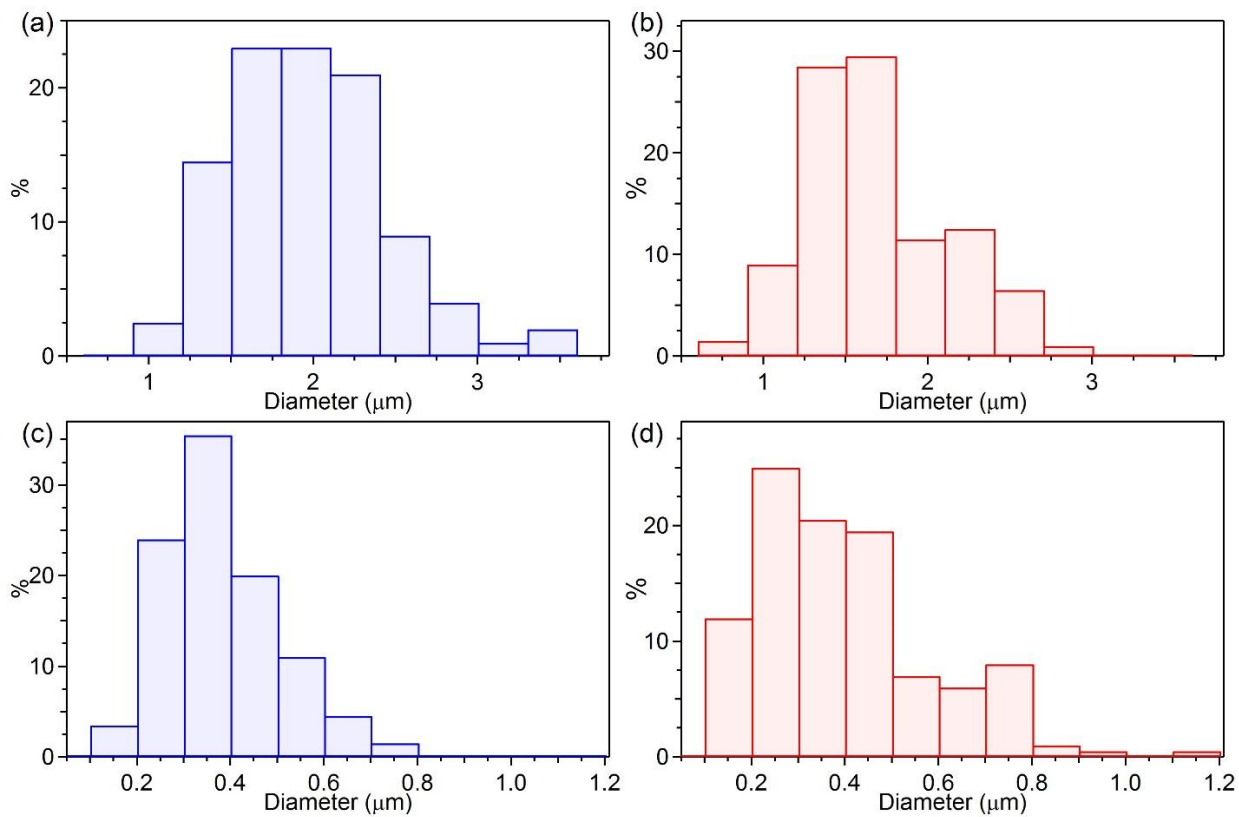

Figure S1. Diameter distributions of fibers based on F8BT (a, b) and MEH-PPV/PVP (c, d) electrospun in controlled nitrogen atmosphere (a, c) and in air (b, d).

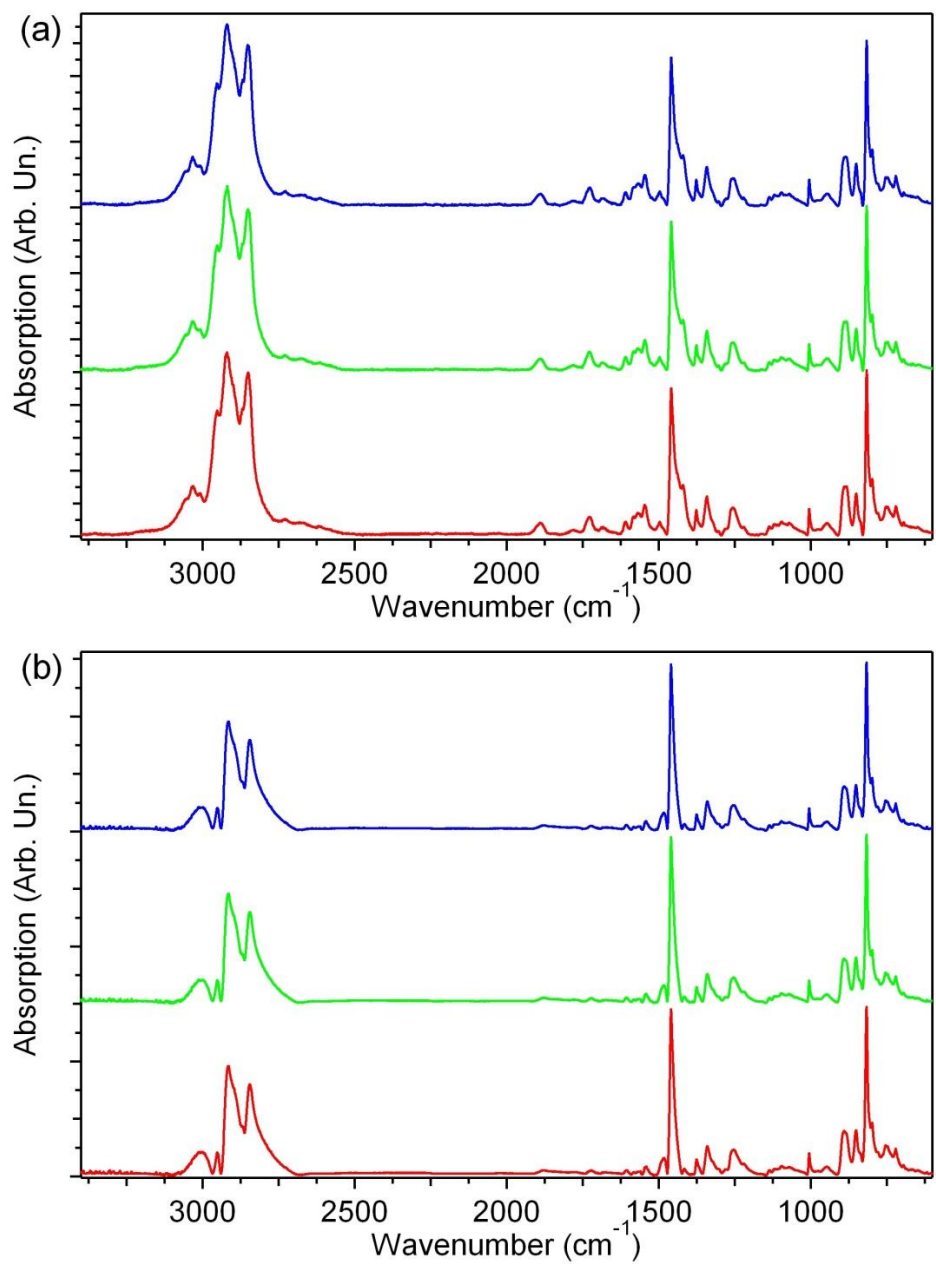

Figure S2. (a,b) FTIR absorption spectra of free-standing mats of F8BT fibers realized in controlled atmosphere (a) and in air (b). Red, green and blue lines correspond to spectra acquired before and after 30 minutes and 60 minutes of UV irradiation, respectively. Spectra are vertically shifted for better clarity.

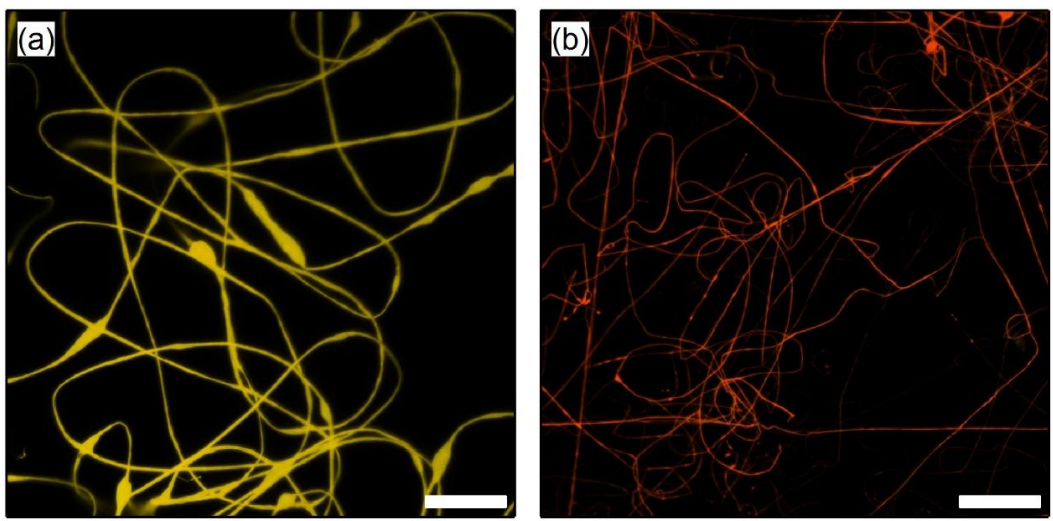

Figure S3. Fluorescence confocal micrographs of F8BT (a) and MEH-PPV/PVP (b) fibers electrospun in air. Scale bars = 100 μm. Excitation wavelengths: 408 nm (a) and 488 nm (b).

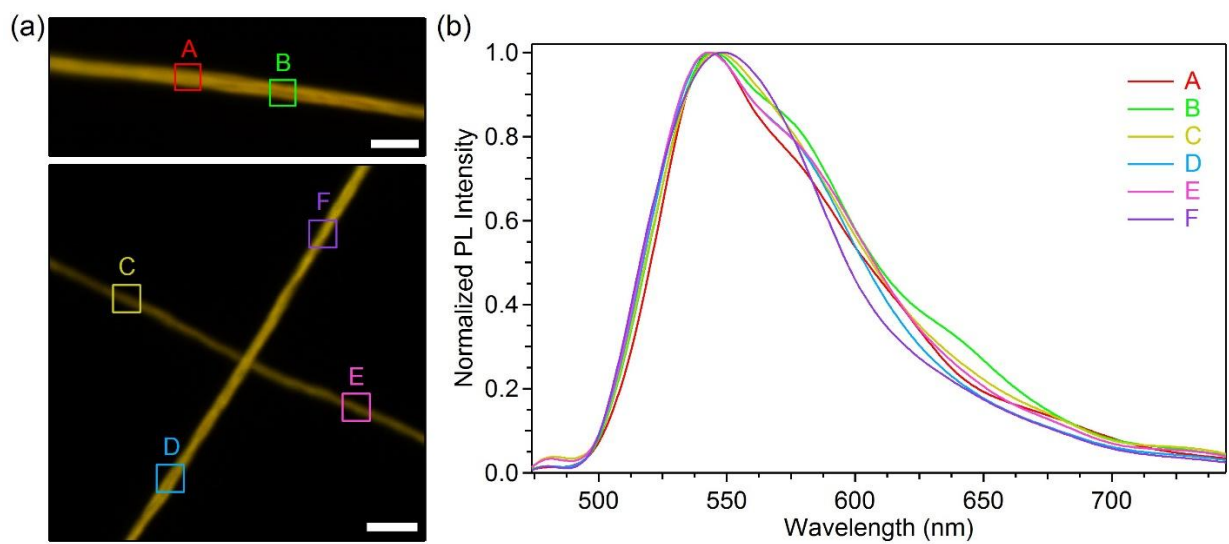

Figure S4. Confocal fluorescence images (a) and spatially-resolved photoluminescence spectra (b) of F8BT fibers electrospun in air. Scale bar = 5 μm. Excitation wavelength: 408 nm. Each emission spectrum, measured in a different region of the fibers (squares in a), is normalized to its maximum value. The size of each analyzed square is 2.5×2.5 μm$^2$.

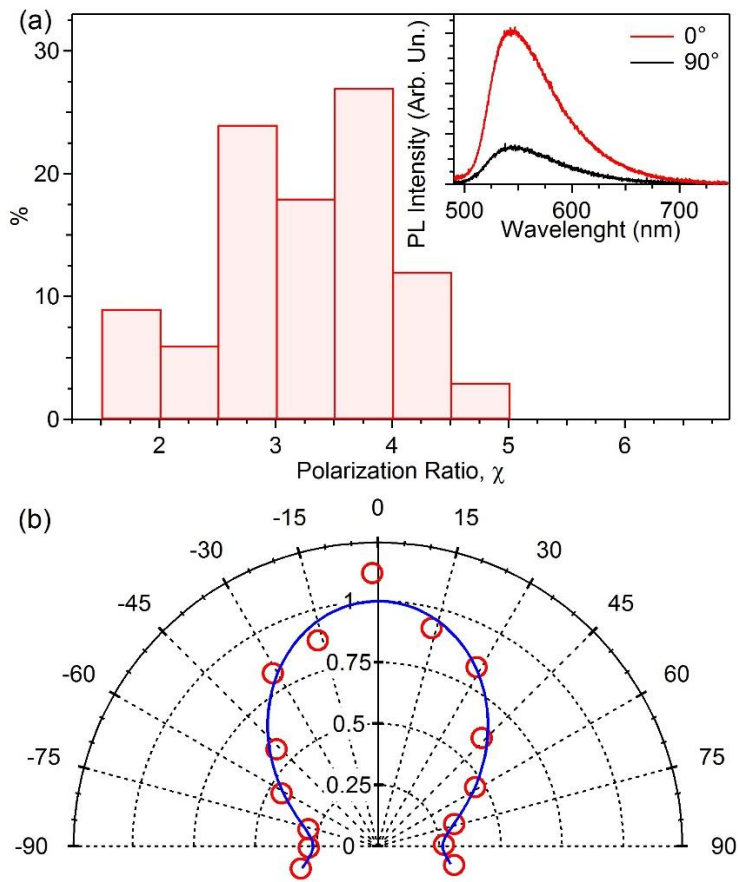

Figure S5. (a) Distribution of photoluminescence polarization ratio for single F8BT fibers electrospun in air. Inset: corresponding polarized emission spectra obtained with the analyzer axis parallel (red line) and perpendicular to (black line) the fiber length. (b) Normalized photoluminescence intensity (circles) vs. polarization angle of the analyzer filter, θ, measured with respect to the fiber longitudinal axis. Continuous line: best fit to data by a Malus law.